# The KICK-OFF of 6G Research Worldwide: An Overview

Wei Jiang[1,2], and Hans D. Schotten[2,1]
[1]*German Research Centre for Artificial Intelligence (DFKI),* Germany
[2]*Institute for Wireless Communication and Navigation, Technische University (TU) of Kaiserslautern,* Germany
wei.jiang@dfki.de

*Abstract*—The fifth-generation (5G) mobile system is now being deployed across the world and the scale of 5G subscribers is growing quickly in many countries. The attention of academia and industry is increasingly shifting towards the sixth generation (6G) and many pioneering works are being kicked off, indicating an important milestone in the history of 6G. At this juncture, an overview of the current state of the art of 6G research and a vision of future communications are of great interest. This paper thus investigates up-to-date 6G research programs, ambitions, and main viewpoints of representative countries, institutions and companies worldwide. Then, the key technologies are outlined and a vision on ``What 6G may look like?'' is provided. This paper aims to serve as an enlightening guideline for interested researchers to quickly get an overview when kicking off their 6G research.

## I. INTRODUCTION

The era of the fifth-generation (5G) mobile communications arrived in April 2019, when South Korea's three mobile operators - SK Telecom, LG U+, and KT - were competing with the United State's carrier Verizon to launch the world's first 5G network. In the past two years, we have witnessed a strong expansion of 5G networks across the world and a great growth of 5G subscriptions in major countries. At the end of 2020, for instance, the penetration rate of 5G usage in South Korea had surpassed 15.5% while China had deployed more than 700,000 base stations to serve around 200 million 5G subscribers. Keeping up with the tradition that a new generation of mobile system is developed every decade, it is time for academia and industry to shift their focus towards the next generation.

Although a debate is ongoing on whether we need the sixth generation (6G) or whether counting should be terminated at 5G, like the Microsoft's approach to use Windows 10 as the ultimate version, some pioneering programs have been initiated as early as 2018. In April, the University of Oulu in Finland announced the world's first 6G research project called *6G-Enabled Wireless Smart Society and Ecosystem (6Genesis)*, as part of Academy of Finland's flagship program. In July, the International Telecommunication Union Telecommunication (ITU-T) standardization sector built a focus group named *Technologies for Network 2030* to study the capabilities of networks dedicated to 2030 and beyond. In 2019, the European Commission's Horizon 2020 program started a call *ICT-20 5G Long Term Evolution* for beyond 5G research. The European Commission has also announced its strategy to accelerate investments in Europe's *Gigabit Connectivity* including 5G and 6G to shape Europe's digital future. Other major players in mobile communications, such as the United States, China, Germany, Japan, and South Korea, have already initiated their national 6G research officially or at least announced their ambitions.

At this junction, an overview of the current state of the art of on-going 6G research programs and a vision of future communications are of interest. In September 2018, David and Berndt published the world's first 6G article [1], which reviews the key services and innovations from the first-generation mobile system to 5G and debates on whether we need beyond 5G. Since then, two dozens of papers focusing on vision, use cases, and key technologies of 6G have been published in the literature. Some representative works are summarized in Tab. I. However, none of these papers keeps an eye on current 6G research campaign and an investigation on the state of the art is missing. To memorize the first milestone (kicking off) in the history of 6G, this paper aims to provide an overview of 6G research programs, ambitions, and main viewpoints of representative countries, institutions, and companies. Then, the promising technologies are outlined and a vision on "*What 6G may look like?*" is presented. This paper aims to serve as an enlightening guideline for interested researchers to quickly get an overview when they kick off their 6G research.

The rest of this paper is organized as follows: the next section focuses on the research campaign in Europe, Sec. III on other regions of the world, and Sec. IV on standardization bodies and industry. Finally, Sec. V concludes this paper by outlining the promising technologies and providing a vision.

TABLE I. STATE-OF-THE-ART 6G PUBLICATIONS

| Subject | | Reference |
|---|---|---|
| Vision | | [1]-[3] |
| Use Case | | [4]-[6] |
| Survey | | [6]-[9] |
| Key Technology | AI/ML | [7], [10]-[12] |
| | THz | [13]-[15] |
| | VLC | [7], [8], [16] |
| | NTN | [6], [7], [17] |
| | Blockchain | [6], [8], [18] |
| | others | [19], [20] |

## II. 6G Research in Europe

We emphasize the 6G research in Europe because their programs are not only the earliest worldwide but also the most transparent and open with rich public-available information. In the past decade, Europe has successfully carried out their 5G research and development under the organization of the 5G Infrastructure Public Private Partnership (5G-PPP). Some key 5G concepts and technologies such as network function virtualization (NFV) and Mobile Edge Computing (MEC) also known as Multi-access Edge Computing were proposed and standardized by European Telecommunications Standards Institute (ETSI). In April 2018, the University of Oulu in Finland announced the world's first 6G research project called 6G-Enabled Wireless Smart Society and Ecosystem (6Genesis) as part of Academy of Finland's flagship program. It focuses on ground-breaking 6G research with four interrelated strategic areas including wireless connectivity, distributed computing, devices and circuit technology, and services and applications. After publishing the world's first 6G White Paper in September 2019 as an outcome of the first 6G Wireless Summit, a series of white papers covering 12 areas of interest such as machine learning, edge intelligence, localization, sensing, and security have been published. Within its eighth Framework Programme (FP8) for Research and Technological Development also known as *Horizon 2020*, the European Commission started beyond 5G research through ICT-20-2019 call *5G Long Term Evolution*. Eight pioneering projects were kicked off in early 2020 after a competitive selection process with a total of 66 high-quality proposals submitted by consortia consisting of vendors, mobile operators, academia, research institutions, small companies, and verticals. At the beginning of 2021, another batch of research projects focusing on 6G sponsored through its call ICT-52-2020 *Smart Connectivity beyond 5G* were kicked off. The details of ICT-20 and ICT-52 research projects are summarized in Table II. Following the success of Horizon 2020 5G-PPP program, the research and development of 6G in Europe will be continued with the upcoming Public Private Partnership (PPP) *Smart Network & Services* under the ninth Framework Program (FP9) also called Horizon Europe. In February 2020, the European Commission has also announced its strategy to accelerate investments in Europe's *Gigabit Connectivity* including 5G and 6G to shape Europe's digital future.

## III. 6G Research Across the World

Through 5G, nations recognized the importance of mobile systems for economic prosperity and national security. In the past year, many countries have announced ambitious plans on the development of 6G or already launched research initiatives officially.

### A. The United States

In 2016, the U.S. Defense Advanced Research Projects Agency (DARPA), along with companies from the semiconductor and defense industries like Intel, Micron, and Analog Devices, have set up the joint university microelectronic project (JUMP) with six research centers to address existing and emerging challenges in microelectronic technologies. *The Center for Converged TeraHertz Communications and Sensing (ComSenTer)* aims to develop technologies for a future cellular infrastructure. The Federal Communications Commission (FCC) announced in March 2019 that it opens experimental license for the use of frequencies between 95GHz and 3THz for 6G and beyond, fostering the test of THz communications. In October 2020, the Alliance for Telecommunications Industry Solutions (ATIS) launched *Next G Alliance* with founding members include AT&T, T-Mobile, Verizon, Qualcomm, Ericsson, Nokia, Apple, Google, Facebook, Microsoft, etc. It is an industry initiative that intends to advance North American mobile technology leadership in 6G over the next decade through private sector-led efforts. With a strong emphasis on technology commercialization, its ambition is to encompass the full lifecycle of 6G research and development, manufacturing, standardization, and market readiness. Another U.S. company, SpaceX, which is famous for its revolutionary innovation of reusable rockets, announced the Starlink project in 2015. Starlink is a very large-scale LEO communication satellite constellation aiming to offer ubiquitous Internet access services across the whole planet. The FCC has approved its first-stage plan to launch 12,000 satellites and another application for 30,000 additional satellites is under consideration. Since its first launch in May 2019, more than 1100 satellites have been successfully deployed through 19 times space launching by far. In the first two months of 2021, around 250 Starlink satellites have been deployed through 4 times launch, reaching its previous plan of deploying 120 satellites per month through two times launch. Currently, the Starlink service is under public testing in North America and UK and is expected to offer commercial service the mid of this year. It is too exaggerated when somebody claims that Starlink will replace 5G or it will be 6G, but the impact of such a very-large-scale LEO satellite constellation on 6G should be seriously taken into account in mobile industry.

### B. China

As early as 2018, the 5G working group at the Ministry of Industry and Information Technology of China started a concept study of potential 6G technologies, making China one of the first countries to explore 6G technology. In November 2019, the Ministry of Science and Technology of China has officially kicked off the 6G technology research and development works coordinated by the ministry, together with five other ministries or national institutions. A promotion working group from government that is in charge of management and coordination, and an overall expert group that is composed of 37 experts from universities, research institutes, and industry were established at this event. In November 2020, the first 6G experimental satellite developed by the University of Electronic Science and Technology of China was successfully launched. Its task is to test communications from space using high-frequency terahertz spectrum. The leading Chinese ICT company, Huawei, said the company is now at the initial stage of 6G research. It provided a roadmap of 6G development with major milestones, including the vision for 6G by around 2023, standardization by 2026, rolling-out relevant technologies by 2028, and preliminary commercial deployment by 2030. In 2020, another telecommunication equipment giant ZTE and one of the three Chinese mobile operators China Unicom initiated their cooperation on 6G technological innovation and standards.

TABLE II. SUMMARY OF EU BEYOND-5G AND 6G RESEARCH PROJECTS

| Call | Project | Main Contribution |
|---|---|---|
| ICT-20 — 5G Long Term Evolution | 5G-CLARITY | To develop a beyond 5G architecture for private networks, which provide enhanced communication services through a novel access network integrating 5G, Wi-Fi, Li-Fi (Light-Fidelity), computing, and transport resources. AI-driven network automation featuring SDN/NFV components and an AI engine is developed. |
| | 5G-COMPLETE | It intends to *revolutionize the 5G architecture* by means of merging communication, computing, and storage resources efficiently over a unified ultra-high-capacity fiber-wireless access network. It converges the fronthaul, midhaul, and backhaul of 5G New Radio (NR) into one common Ethernet-based platform and transforms the radio network into a low-power distributed computer. |
| | 5G-ZORRO | This project seeks to implement zero-touch security and trust for ubiquitous computing and connectivity in evolved 5G networks. Distributed AI is adopted to realize automated end-to-end network management across multiple operators and providers with minimal manual intervention (zero-touch network automation). Distributed ledger technology is employed to support flexible and efficient distributed security and trust across the various parties. |
| | ARIADNE | The aim of ARIADNE is to design an intelligent communication system to address the challenges raised by the adoption of higher frequency bands in *beyond 5G*. Novel radio transmission schemes using D-Band spectral resources over 100GHz and advanced connectivity based on meta-surfaces, which can tune the reflectors to shape the propagation characteristics, are developed. AI-driven processing and network management is considered to dynamically configure high-frequency communication and meta-surfaces. |
| | INSPIRE-5G+ | This project intends to advance intelligent security and pervasive trust for *5G and Beyond* networks taking advantage of novel approaches and solutions empowered by AI, ML, and blockchain technologies such as proactive security while being trustworthy. It also aims to deliver intelligent and trusted multi-tenancy, improve the control of systems, and lower the vulnerability and compromise of infrastructure providers and tenants. |
| | LOCUS | The goal of this project is to further advance the 5G system by means of offering accurate localization function and inferring features and behavioral patterns from raw location data. It is expected to substantially improve the functionality of *the future network* through on-demand, network-native localization-as-a-service for ubiquitous vertical applications. |
| | MonB5G | It towards delivering zero-touch automation for distributed management and orchestration of massive-scale network slices with diverse performance requirements in *beyond 5G*. With cutting-edge AI-based monitoring, analysis, and decision-making approaches, it develops a hierarchical, fault-tolerant, automatic management system that improves energy efficiency, security, and trustworthy tailed to multi-stakeholder operation environment. |
| | TERAWAY | It focuses on the development of a mutli-channel, photonics-based THz transceiver that can amplify wireless signals in the optical domain and support multi-beam optical beamforming with high directivity. Applying SDN-based radio resources management, it provides the capability of flexibly utilizing a pool of an ultra-wide range of spectral resources covering W, D, and THz bands. |
| ICT-52 — Smart Connectivity beyond 5G | 6G BRAINS | It seeks to apply multi-agent deep reinforcement learning to implement cross-layer resource allocation to support massive connection in *beyond 5G and 6G networks*. Four key technologies are considered, including highly dynamic device-to-device cell-free networking, disruptive spectrum usage with THz and optical wireless communications (OWC), end-to-end intelligent network architecture, and AI-enhanced high-resolution 3D simultaneous localization and mapping (SLAM). |
| | AI@EDGE | This project commits to two complementary paradigms in *beyond 5G system*, i.e., (i) AI-for-networks: closed-loop network automation offering flexible and programmable life-cycle management of secure, reusable, and trustworthy AI models; and (ii) network-for-AI: a connect-compute platform for instantiating and managing resilient, secure end-to-end network slices supporting a wide variety of AI services and applications. |
| | DAEMON | It focuses on network intelligence (NI) that provides fully autonomous management of *beyond 5G networks*. Through designing an end-to-end NI-native architecture and specific-purpose AI models, it is expected to achieve extremely high performance, more efficient use of radio, computing, and energy resources, and extremely high reliability beyond that of 5G systems. |
| | DEDICAT 6G | The purpose of DEDICAT 6G is to apply AI and blockchain techniques to develop a smart connectivity platform that enables dynamic coverage extension by exploiting novel mobile terminals and provides distributed intelligence for human-centric real-time experience with assured security, privacy and trust in *6G wireless networks*. |
| | Hexa-X | It is a flagship project for *beyond 5G/6G* vision and intelligent fabric of technology enablers, which include novel radio access technologies at high frequency bands, high-resolution localization and sensing, connected intelligence though AI-driven air interface and governance for future networks, and 6G architectural enablers for network disaggregation and dynamic dependability. |
| | MARSAL | A comprehensive framework for management and orchestration of network resources in *5G and beyond systems* is developed by applying converged fixed-mobile, optical-wireless infrastructure, distributed cell-free and serial fronthaul approaches, and secure multi-tenant slicing based on AI and blockchain. |
| | REINDEER | Going beyond the concepts of large-scale intelligent surfaces and cell-free networking, this project aims to develop a smart connect-compute platform consisting of a fabric of distributed communication, computing, and storage resources. With the proposed RadioWeaves technology, it brings a large number of antennas and intelligence close to devices, providing quasi-infinite capacity, zero-perceived latency, and more efficient usage of spectral, computational, and energy resources. |
| | RISE-6G | This project studies reconfigurable intelligent surfaces (RIS) technology to realize controllable, programmable radio propagation environment that goes well *beyond the 5G capabilities*. Novel multi-RIS network architecture and operation, fundamental limits analysis on realistic radio propagation models, and dynamically online configuration in terms of high-capacity connectivity, energy efficiency, EMF exposure, and localization accuracy are investigated. |
| | TeraFlow | It aims to design a novel cloud-native architecture for *beyond 5G networks* that will substantially advance the state of the art. A novel SDN controller compatible with current NFV and MEC frameworks is developed to provide revolutionary features for both flow management (service layer) and optical-microwave integration (infrastructure layer), while incorporating ML-based security and forensic evidence for multi-tenancy based on distributed ledger technologies. |

## C. Japan

In late 2017, a working group was established by the Japanese Ministry of Internal Affairs and Communications to study next-generation wireless technologies. Their findings are that 6G should provide transmission rates at least 10 times faster than 5G, near-instant connection, and massive connection of 10 million devices per square kilometer. In early 2020, the Japanese government set up a dedicated panel including representatives from the private sector and academia to discuss technological development, potential use cases, and policy. Japan reportedly intends to dedicate around $2 billion to encourage private-sector research and development for 6G technology.

## D. South Korea

South Korea announced a plan to set up the first 6G trial in 2026 and is expected to spend around $169 million over five years to develop key 6G technologies. The trial aims to realize the peak data rate of 1Tbps and extreme-low latency that is 10 times lower than that of 5G. The government of South Korea will push the research programs in six key areas (hyper-performance, hyper-bandwidth, hyper-precision, hyper-space, hyper-intelligence, and hyper-trust) to preemptively secure next-generation technology. As Huawei's 5G technology is banned in the USA, Australia and the UK, Samsung Electronics speeds up their global ambition to become a major telecommunication vendor. Samsung Research has established the Next-Generation Communication Research Center on the basis of Samsung Research's Standard Research Team that carries out research on 6G, as the largest among Samsung Research's units. LG Electronics announced its ambition to lead global standardization on 6G and create new business opportunities. As a follow-up, it launched a 6G Research Center in January 2019

together with the Korea Advanced Institute of Science and Technology (KAIST).

## IV. STANDARDIZATION BODIES AND INDUSTRY

### A. ITU

As a pointer to the new horizon for the future digital society and networks, ITU-T *Focus Group Technologies for Network 2030* (FG NET-2030) was established as early as July 2018 to identify the gaps and technological challenges towards the capabilities of networks for the year 2030 and beyond, when it is expected to satisfy extreme performance requirements to support disruptive use cases such as holographic-type communications, Tactile Internet, multi-sense networks, and digital twin. Although it mainly focuses on fixed communication networks, the future network architecture, requirements, use cases, and capabilities of the networks identified in this group will be a guideline for the definition of the 6G mobile system. The ITU Radiocommunication Sector (ITU-R) has recently published Recommendation ITU-R M.2150 titled 'Detailed specifications of the radio interfaces of International Mobile Telecommunications (IMT)-2020', which can be regarded as the finalization of 5G standardization. Considering the successful accomplishments by ITU for the evolution of IMT-2000 (the third generation or 3G), IMT-Advanced (the fourth generation or 4G), and IMT-2020 (5G), the similar process will be applied once again for the development of *IMT towards 2030 and beyond*. According to the IMT process, ITU-R starts the study of ITU-R Vision on 6G as the first step and then publishes the minimum requirements and evaluation criteria for *IMT towards 2030 and beyond* in the middle of the 2020s and will step into invitation for proposals and the evaluation phase afterwards. At its meeting in February 2020, ITU-R working party 5D decided to start the study on future technology trends and plans to complete this study at the meeting in June 2022. It invited organizations within and external to the ITU-R to provide inputs for its June and October meetings in 2021, which will help the development of the first draft "Future Technology Trends towards 2030 and beyond". ITU-R is also responsible for organizing the world radiocommunication conference (WRC) that governs the frequency assignment, being hold every three to four years. In WRC-19 held in 2019, the spectrum allocation issue for the 5G system was determined. It is expected that the WRC probably scheduled in 2023 (WRC-23) will discuss the spectrum issues for 6G and the spectrum allocation for 6G communications may be formally decided in 2017 (WRC-27).

### B. 3GPP

In early 2019, the third-generation partnership project (3GPP) has frozen the Release 15 (Rel. 15 or R15) specifications, as the first phase of 5G standards. In July 2020, the subsequent release (i.e., Rel. 16) has been completed as the second phase of 5G standards [21]. Currently, a more advanced version (Rel. 17) is being standardized by 3GPP and is expected to be completed in the early of 2021, in spite of a delay due to the COVID-19 pandemic. Driven by a multitude of key stakeholders from the traditional mobile industry, a wide range of verticals, and the non-terrestrial industry, it is envisioned as the most versatile release in 3GPP history in terms of features content, including NR over non-terrestrial networks, NR beyond 52.6GHz, NR sidelink enhancement, network automation, etc. 3GPP is expected to standardize several subsequent releases to further evolve the 5G system, which can be called 5G+ or 5G Evolution. According to the experiences got in previous generations, 6G will be a disruptive system that cannot be developed following such a backward-compatible manner. In parallel, therefore, 3GPP is expected to initiate the study item for 6G around the year 2025, followed by the phase of specification, to guarantee the first commercial deployment roll-out of 6G by 2030.

### C. Industry

In the mobile industry, the vendors, mobile operators, and device manufacturers are the real driving forces. Hence, the viewpoints of the industry, especially the major vendors play an important role in the development of new generation. In January 2020, NTT Docomo published its white paper "5G Evolution and 6G", followed by other major players. We provide a summary of their main viewpoints in Tab. III.

TABLE III. SUMMARY OF 6G WHITE PAPERS FROM INDUSTRY

| Company | White Paper | Release | Main Viewpoint |
|---|---|---|---|
| NTT Docomo | 5G Evolution and 6G | Jan. 2020 | 5G is the first generation mobile system utilizing the mmWave band and further improvement on coverage and mobility of mmWave technology is needed for 5G evolution. Extreme-high capacity and extreme-low latency in uplink from physical to cyber space, while extreme-high reliability and extreme-low latency in downlink from cyber to physical space, are needed to support the cyber-physical fusion, requiring substantial enhancement of uplink transmission in 5G evolution and 6G. The requirements of 6G will be new combinations of 5G requirements for new use cases or extreme requirements for specific use cases. |
| Nokia | Communications in the 6G era | Mar. 2020 | 5G will continue to be evolved to improve the performance by adopting new technologies in a backward-compatible manner. Modifications that are not compatible with the 5G framework or can only be implemented with high cost will be part of 6G, which provides the inter-connection of physical, biological, and digital worlds. Key 6G technologies are likely to be high spectrum bands and cognitive spectrum sharing, AI-driven air interface, new networking paradigms involving private networks and RAN-Core convergence, the integration of communication, sensing, and localization, extreme latency and reliability, and new security and privacy. |
| Rohde&Schwarz | 5G evolution - on the path to 6G | Mar. 2020 | It is important to follow standardization activities to get insight into what vendors, mobile operators, and device makers are attempting to accomplish and fix. Weaknesses in the 5G standard are the driver on the path towards 6G, and this is especially true with respect to the weaknesses in security and privacy. What will distinguish 6G development is that nations now recognize the significance of wireless standards for economic prosperity and national security. |
| Samsung | The next hyper-connected experience for all | Jul. 2020 | The mega-trends driving the mobile industry towards 6G are massive connected machines, AI, openness of networks, and increasing societal impact of mobile communications. 6G will support three key services: truly immersive extended reality (XR), high-fidelity mobile hologram, and digital twin. Key technological enablers are THz technologies, advanced antenna techniques, evolved duplex technology, dynamic network topology, spectrum sharing, comprehensive AI, split computing, and high-precision network. |
| Ericsson | Ever-present intelligent communication: a research outlook towards 6G | Nov. 2020 | Despite the built-in flexibility of 5G, it will evolve to 6G following the pull from society needs and the push from more advanced technologies. The main driving forces behind 6G are trustworthiness, sustainability, automatization, digitalization, and limitless connectivity anywhere, anytime, and for anything. Going beyond connectivity, 6G should become a trusted compute and data platform encouraging innovation and serving as the information backbone of society. |
| Huawei | Defining 5.5G for a Better, Intelligent World | Nov. 2020 | Continuous evolution is necessary to fully unleash the potential of the technology, and the industry should work together to drive a thriving 5.5G ecosystem by making the most of sub-100 GHz spectrum. Three new scenarios, i.e., Uplink Centric Broadband Communication (UCBC) facilitating high-definition video uploading and machine vision, Real-Time Broadband Communication (RTBC) offering an immersive experience, and Harmonized Communication and Sensing (HCS) for autonomous driving and connected drones, are envisaged. |

TABLE IV. POTENTIAL KEY 6G TECHNOLOGIES

| Category | Key Technologies | Notes |
|---|---|---|
| Spectrum | mmWave | 5G is the first mmWave mobile system and mobility/coverage of sub-100GHz need to enhance in 6G. |
| Spectrum | THz | Offering abundant spectral resources but many obstacles on hardware implementation (RF, antenna) should be solved. |
| Spectrum | VLC | Almost-unlimited, licence-free bandwidth in 400~800THz, low-cost implementation with LED, no electromagnetic interference |
| Spectrum | OWC | It uses optical bands (infrared, visible light, and ultraviolet), and free-space optics using Laser has special applications. |
| Antenna | RIS | It can create a small, controllable, and programmable radio propagation environment as a new degree of freedom. |
| Antenna | OAM | To achieve higher data rate by applying orthogonal orbital angular momentum (OAM) modes to multiplexing multiple signals. |
| Antenna | Lens Antenna | Highly-directive antennas for transmission over high frequency band, implemented by transformation optics or metasurfaces. |
| Air Interface | mMIMO [22]-[30] | The number of elements in an antenna array is further increased to an ultra level to support extreme performance requirements. |
| Air Interface | Holographic Radio | A new method to create a spatially-continuous electromagnetic aperture to enable more spatial multiplexing than massive MIMO. |
| Air Interface | CF-mMIMO | It provides a novel method to eliminate inter-cell interference and improve user-experienced data rate at cell edge [31]-[33]. |
| Air Interface | NOMA | An efficient multi-access method [34]-[36] making use of near-far effect, supporting higher multi-user capacity and massive connection. |
| Networking | RAN Slicing | An extension of network slicing to improve the efficiency of RAN resources via virtualization, softwarization, and cloudification. |
| Networking | O-RAN | It aims to bring cost-efficient, resource-efficient RAN architecture by introducing openness and intelligence. |
| Networking | Automation | Using AI and Big Data analytics to automate network management [37]-[42] to improve performance and lower expenditure. |
| Networking | Security/Privacy | Innovative technologies are needed to guarantee security and privacy of users and vertical industries in the post-quantum era. |
| NTN | GEO satellite | It can provide ubiquitous global coverage, but its throughput, cost, and transmission latency are concerns. |
| NTN | LEO satellite | The impact of large-scale LEO constellation on 6G should be seriously considered with the rise of revolutionized space technology. |
| NTN | HAP [43], [44] | At the altitude of 20km, it provides cost-efficient, flexible deployment with large coverage for new applications. |
| NTN | UAV | It enables a wide variety of new use cases and new applications but bringing disruptive technological challenges. |
| Paradigm | AI [45]-[50] | Deep learning, federated/transfer learning are highly potential for the realization of AI-for-Network and Network-for-AI. |
| Paradigm | Blockchain | It has unique features such as immutability, decentralization, transparency, and trustworthiness. |
| Paradigm | QC | It might offer absolute security for communication networks and shows great advances towards practical deployment recently. |
| Paradigm | Edge intelligence | Provide pervasive intelligence using edge computing resources for low-latency, computing-intensive services. |
| Paradigm | Split Computing | A novel resource management method to efficiently utilize the computing resources across terminals, edge, and cloud. |
| Paradigm | CCCS | Convergence of communication, computing, and storage resources with the integration of sensing and localization functionalities. |

## V. CONCLUSION REMARKS: KEY TECHNOLOGIES AND VISION

Absorbing the ingredients of 6G scientific publications, industrial white papers, among all others, we identify potential 6G key technologies, as listed in Tab. IV, and provide a vision on "*What 6G might look like*?" as follows. It is envisaged that 6G will be an unprecedented generation and remarkably differentiated with the previous generations:

- It will be a radio-optical system operating across radio frequency bands of sub-6GHz, mmWave, and THz to optical bands using infrared, visible lights, and ultraviolet waves.
- It will become an intelligent network maximizing the synergy of AI technologies and mobile communications.
- It will be a space-aerial-terrestrial network integrating terrestrial and non-terrestrial networks to provide ubiquitous 3D coverage of the whole surface of the Earth.
- It will be a compute-connect platform where the convergence of distributed communication, computing, storage, sensing, localization, and control are well achieved.

REFERENCES

[1] K. David and H. Berndt, "6G vision and requirements: Is there any need for beyond 5G?"IEEE Veh. Technol. Mag., vol. 13, no. 3, pp. 72–80, Sep.2018.
[2] W. Saad et al., "A vision of 6G wireless systems: Applications, trends, technologies, and open research problems,"IEEE Netw., vol. 34, no. 3, pp.134–142, May 2020.
[3] F. Tariqet et al., "A speculative study on 6G,"IEEE Wireless Commun., vol. 27, no. 4, pp. 118–125, Aug. 2020.
[4] G. Liuet et al., "Vision, requirements and network architecture of 6G mobile network beyond 2030,"China Commun., vol. 17, no. 9, pp. 92–104, Sep.2020.
[5] M. Giordani et al., "Toward 6G networks: Use cases and technologies,"IEEE Commun. Mag., vol. 58, no. 3, pp. 55–61, Mar. 2020.
[6] W. Jiang et al., "The road towards 6G: A comprehensive survey," IEEE Open J. Commun. Society, vol. 2, pp. 334 – 366, Feb. 2021.

[7] E. C. Strinati et al., "6G: The next frontier: From holographic messaging to artificial intelligence using subterahertz and visible light communication," IEEE Veh. Technol. Mag., vol. 14, no. 3, pp. 42–50, Sep. 2019.

[8] T. Huang et al., "A survey on green 6G network: Architecture and technologies," IEEE Access, vol. 7, pp. 175 758–175 768, Dec. 2019.

[9] H. Viswanathan and P. E. Mogensen, "Communications in the 6G era," IEEE Access, vol. 8, pp. 57 063–57 074, Mar. 2020.

[10] K. B. Letaief et al., "The roadmap to 6G: AI empowered wireless networks," IEEE Commun. Mag., vol. 57, no. 8, pp. 84–90, Aug. 2019.

[11] W. Jiang and F.-L. Luo, "Editorial: Special topic on computational radio intelligence: One key for 6G wireless," ZTE Communications, vol. 17, no. 4, pp. 1–3, Dec. 2019.

[12] N. Kato et al., "Ten challenges in advancing machine learning technologies toward 6G," IEEE Wireless Commun., vol. 27, no. 3, pp. 96–103, Jun. 2020.

[13] T. S. Rappaport et al., "Wireless communications and applications above 100 GHz: Opportunities and challenges for 6G and beyond," IEEE Access, vol. 7, pp. 78 729–78 757, Jun. 2019.

[14] K. Rikkinen et al., "THz radio communication: Link budget analysis toward 6G," IEEE Commun. Mag., vol. 58, no. 11, pp. 22–27, Nov. 2020.

[15] M. Polese et al., "Toward end-to-end, full-stack 6G terahertz networks," IEEE Commun. Mag., vol. 58, no. 11, pp. 48–54, Nov. 2020.

[16] N. Chi et al., "Visible light communication in 6G: Advances, challenges, and prospects," IEEE Veh. Technol. Mag., vol. 15, no. 4, pp. 93–102, Dec.2020.

[17] M. Kishk et al., "Aerial base station deployment in 6G cellular networks using tethered drones: The mobility and endurance tradeoff," IEEE Veh. Technol. Mag., vol. 15, no. 4, pp. 103–111, Dec. 2020.

[18] M. Z. Chowdhury et al., "6G wireless communication systems: Applications, requirements, technologies, challenges, and research directions," IEEE Open J. Commun. Society, vol. 1, pp. 957–975, Jul. 2020.

[19] C. Huang et al., "Holographic MIMO surfaces for 6G wireless networks: Opportunities, challenges, and trends," IEEE Wireless Commun., vol. 27, no. 5, pp. 118–125, Oct. 2020.

[20] C.-X. Wang et al., "6G wireless channel measurements and models: Trends and challenges," IEEE Veh. Technol. Mag., vol. 15, no. 4, pp. 22–32, Dec.2020.

[21] A. Ghosh et al., "5G evolution: A view on 5G cellular technology beyond 3GPP release 15," IEEE Access, vol. 7, pp. 127 639–127 651, Sep. 2019.

[22] W. Jiang and H. D. Schotten, "Multi-antenna fading channel prediction empowered by artificial intelligence," in Proc. IEEE Veh. Tech. Conf. (VTC), Chicago, USA, Aug. 2018.

[23] W. Jiang and Hans D. Schotten, "A deep learning method to predict fading channel in multi-antenna systems," in Proc. IEEE Veh. Tech. Conf. (VTC), Antwerp, Belgium, May 2020.

[24] W. Jiang and Hans D. Schotten, "Recurrent neural network-based frequency-domain channel prediction for wideband communications," in Proc. IEEE Veh. Tech. Conf. (VTC), Kuala Lumpur, Malaysia, Apr. 2019.

[25] W. Jiang and Hans D. Schotten, "Recurrent neural networks with long short-term memory for fading channel prediction," in Proc. IEEE Veh. Tech. Conf. (VTC), Antwerp, Belgium, May 2020.

[26] W. Jiang and Hans D. Schotten, "Neural networks-based channel prediction and its performance in multi-antenna systems," in Proc. IEEE Veh. Tech. Conf. (VTC), Chicago, UAS, Aug. 2018, pp. 1-6.

[27] W. Jiang, H. Schotten, and J.-Y. Xiang, "Neural network-based wireless channel prediction," in Machine Learning for Future Wireless Communications, F. L. Luo, Eds. United Kindom: John Wiley&Sons and IEEE Press, 2020, ch. 16.

[28] W. Jiang and H. D. Schotten, "A comparison of wireless channel predictors: Artificial intelligence versus Kalman filter," in Proc. IEEE Intl. Conf. on Commun. (ICC), Shanghai, China, May 2019, pp. 1 – 6.

[29] W. Jiang et al., "Long-range MIMO channel prediction using recurrent neural network," in Proc. IEEE Consumer Commun. And Netw. Conf. (CCNC), Las Vegas, USA, Jan. 2020, pp. 1 – 6.

[30] W. Jiang and X. Yang, "An enhanced random beamforming scheme for signal broadcasting in multi-antenna systems," in Proc. IEEE Int. Symp. on Personal, Indoor and Mobile Radio Commun. (PIMRC), Sydney, Austrilia, Sept. 2012, pp. 2055 – 2060.

[31] W. Jiang and Hans D. Schotten, "Cell-free massive MIMO-OFDM transmission over frequency-selective fading channels," IEEE Commun. Lett., vol. 25, no. 8, pp. 2718 – 2722, Aug. 2021.

[32] W. Jiang and H. Schotten, "Impact of channel aging on zero-forcing precoding in cell-free massive MIMO systems," IEEE Commun. Lett., vol. 25, no. 9, pp. 3114 – 3118, Sept. 2021.

[33] W. Jiang et al., "A robust opportunistic relaying strategy for co-operative wireless communications," IEEE Trans. Wireless Commun., vol. 15, no. 4, pp. 2642 – 2655, Apr. 2016.

[34] W. Jiang and T. Kaiser, "From OFDM to FBMC: Principles and Comparisons," in Signal Processing for 5G: Algorithms and Implementations, F. L. Luo and C. Zhang, Eds. United Kindom: John Wiley&Sons and IEEE Press, 2016, ch. 3.

[35] W. Jiang and M. Schellmann, "Suppressing the out-of-band power radiation in multi-carrier systems: A comparative study," in Proc. IEEE Global Commun. Conf. (GLOBECOM), Anaheim, USA, Dec. 2012, pp. 1477 – 1482.

[36] W. Jiang and Z. Zhao, "Low-complexity spectral precoding for rectangularly pulsed OFDM," in Proc. IEEE Veh. Tech. Conf. (VTC), Quebec City, Canada, Sept. 2012, pp. 1 - 5.

[37] W. Jiang et al., "Autonomic network management for software-define and virtualized 5G systems," in Proc. Eur. Wireless, Dresden, Germany, May 2017.

[38] W. Jiang et al., "Experimental results for artificial intelligence-based self-organized 5G networks," in Proc. IEEE Int. Symp. on Personal, Indoor and Mobile Radio Commun. (PIMRC), Montreal, QC, Canada, Oct. 2017.

[39] W. Jiang et al., "Intelligent network management for 5G systems: The SELFNET approach," in Proc. IEEE Eur. Conf. on Net. and Commun. (EUCNC), Oulu, Finland, Jun. 2017, pp. 109–113.

[40] W. Jiang et al., "A SON decision-making framework for intelligent management in 5G mobile networks," in Proc. IEEE Intl. Conf. on Compu. and Commun.(ICCC), Chengdu, China, Dec. 2017.

[41] W. Jiang et al., "Intelligent Slicing: A unified framework to integrate Artificial Intelligence into 5G networks," in Proc. IFIP Wireless and Mobile Netw. Conf. (WMNC), Paris, France, Sept. 2019, pp. 227 - 232.

[42] W. Jiang et al., "An SDN/NFV proof-of-concept test-bed for machine learning-based network management," in Proc. IEEE Intl. Conf. on Compu. and Commun. (ICCC), Chengdu, China, Dec. 2018, pp. 1966 - 1971.

[43] W. Jiang et al., "Opportunistic relaying over aerial-to-terrestrial and device-to-device radio channels," in Proc. IEEE Intl. Conf. on Commun. (ICC), Sydney, Australia, Jul. 2014, pp. 206–211.

[44] W. Jiang et al., "Achieving high reliability in aerial-terrestrial networks: Opportunistic space-time coding," in Proc. IEEE Eur. Conf. on Net. and Commun. (EUCNC), Bologne, Italy, Jun. 2014.

[45] W. Jiang and H. D. Schotten, "Deep learning for fading channel prediction," IEEE Open J. of the Commun. Society, vol. 1, pp. 320–332, Mar. 2020.

[46] W. Jiang et al., "Neural network-based fading channel prediction: A comprehensive overview," IEEE Access, vol. 7, pp. 118 112–118 124, Aug. 2019.

[47] W. Jiang and Hans D. Schotten, "A simple cooperative diversity method based on deep-learning-aided relay selection," IEEE Trans. Veh. Technol., vol. 70, no. 5, pp. 4485 – 4500, May 2021.

[48] W. Jiang and H. Schotten, "Predictive relay selection: A cooperative diversity scheme using deep learning," in Proc. IEEE Wireless Commun. and Netw. Conf. (WCNC), Nanjing, China, Mar. 2021.

[49] W. Jiang and H. Schotten, "Device-to-device based cooperative relaying for 5G network: A comparative review," ZTE Communications, vol. S1, Dec. 2017, pp 60-66.

[50] W. Jiang et al., "A tracking algorithm in RFID reader network," in Proc. Japan-China Joint Workshop on Frontier of Computer Sci. and Techno. (FCST), Fukushima, Japan, Nov. 2006, pp. 164-171.